\documentclass[prl,preprint,showpacs,amsmath,amssymb,superscriptaddress]{revtex4}
\usepackage{graphicx}
\usepackage{amssymb}

\begin{document}

\title{Molecular Beam Epitaxy of LiMnAs}

\author{V. Nov\'{a}k}
\affiliation{Institute of Physics ASCR, v.v.i., Cukrovarnick\'a 10, 162 53 Praha, Czech Republic}
\author{M. Cukr}
\affiliation{Institute of Physics ASCR, v.v.i., Cukrovarnick\'a 10, 162 53 Praha, Czech Republic}
\author{Z. \v{S}ob\'a\v{n}}
\affiliation{Institute of Physics ASCR, v.v.i., Cukrovarnick\'a 10, 162 53 Praha, Czech Republic}
\affiliation{Faculty of Electrical Engineering, Czech Technical University in Prague, Technick\'{a} 2, 166 27 Prague, Czech Republic}
\author{T.~Jungwirth}
\affiliation{Institute of Physics ASCR, v.v.i., Cukrovarnick\'a 10, 162 53 Praha, Czech Republic}
\affiliation{School of Physics and Astronomy, University of Nottingham, Nottingham NG7 2RD, United Kingdom}
\author{X. Mart\'{\i}}
\author{V.~Hol\'{y}}
\author{P.~Horodysk\'a}
\author{P.~N\v{e}mec}
\affiliation{Faculty of Mathematics and Physics, Charles University in Prague, Ke Karlovu 3, 121 16 Prague, Czech Republic}

\begin{abstract}
We report on the molecular beam epitaxy (MBE) growth of high crystalline quality LiMnAs. The introduction of a group-I alkali metal element Li with flux comparable to fluxes of Mn and As has not caused any apparent damage to the MBE system after as many as fifteen growth cycles performed on the system to date.
\end{abstract}


\maketitle




\section{Introduction}

Molecular beam epitaxy (MBE) has proven to be an unrivaled technology for preparing high quality thin-film crystalline materials and layered heterostructures. While it became a standard growth technique in the field of group-IV, III-V and II-VI semiconductors, no epitaxial growth has been reported until recently  \cite{Jungwirth:2010} for compound materials containing group-I elements. Yet, materials of this class exhibit a variety of properties of basic science and application interests. These include semiconducting band structure (e.g. LiZnAs, \cite{Kuriyama}, or LiMnAs, \cite{Jungwirth:2010}) combined with high-temperature antiferromagnetism (e.g. LiMnAs, KMnAs, \cite{Bronger:1986}) or with the predicted ferromagnetism in dilute moment sytems (e.g. LiZnMnAs, \cite{Masek:2007_a}), and with structural compatibility to III-V zinc-blend materials \cite{Achenbach:1981,Jungwirth:2010}. 

Materials from the I-Mn-V group were previously synthesized from elements by high temperature reaction \cite{Bronger:1986,Achenbach:1981}. These materials are  powder-like, polycrystalline, and sensitive to air moisture. LiMnAs, which we focus on in this paper, crystallizes in tetragonal space group with lattice parameters $a=4.26$~\AA\, and $c/a=1.45$, as determined from  x-ray diffraction (XRD) measurements  \cite{Achenbach:1981}. The spin ordering was determined by neutron diffraction to be antiferromagnetic above room temperature \cite{Bronger:1986}. 

The fact that MBE growth of LiMnAs, as well as of other group-I compounds, has remained a virtually unexplored territory is likely a result of the common prejudice that high concentrations of strongly reactive alkali metals might damage the MBE system. An indication that this may not be the case and the first study of LiMnAs epilayers has been recently presented in Ref.~\cite{Jungwirth:2010}. While the report \cite{Jungwirth:2010}  focuses on the semiconducting nature of antiferromagnetic LiMnAs and outlines a general strategy for employing high-temperature antiferromagnetic semiconductors in spintronics, in this paper we provide a more detailed description of the MBE growth of LiMnAs.

\section{MBE growth}

In addition to common sources of arsenic (As$_4$) and manganese, the MBE system (Kryovak) was equipped with an effusion cell containing elemental lithium (in the form of solid Li chunks) in a 5cc pBN crucible. Molecular fluxes of all  three cells
were determined by the beam-flux monitor, taking into account the respective molecular weights and atomic numbers \cite{Parker}. During the growth, the stoichiometric ratio of Li, Mn, and As was adjusted, with a slight excess of As. The corresponding cell temperatures and beam-equivalent pressures are summarized in Tab.~\ref{tab1} for the case of the growth rate of $\approx 0.2\,\mu$m/hour,  used in our experiments.

\begin{figure}[h]
\vspace*{0.3cm}
\begin{center}
\includegraphics[width=0.30\columnwidth]{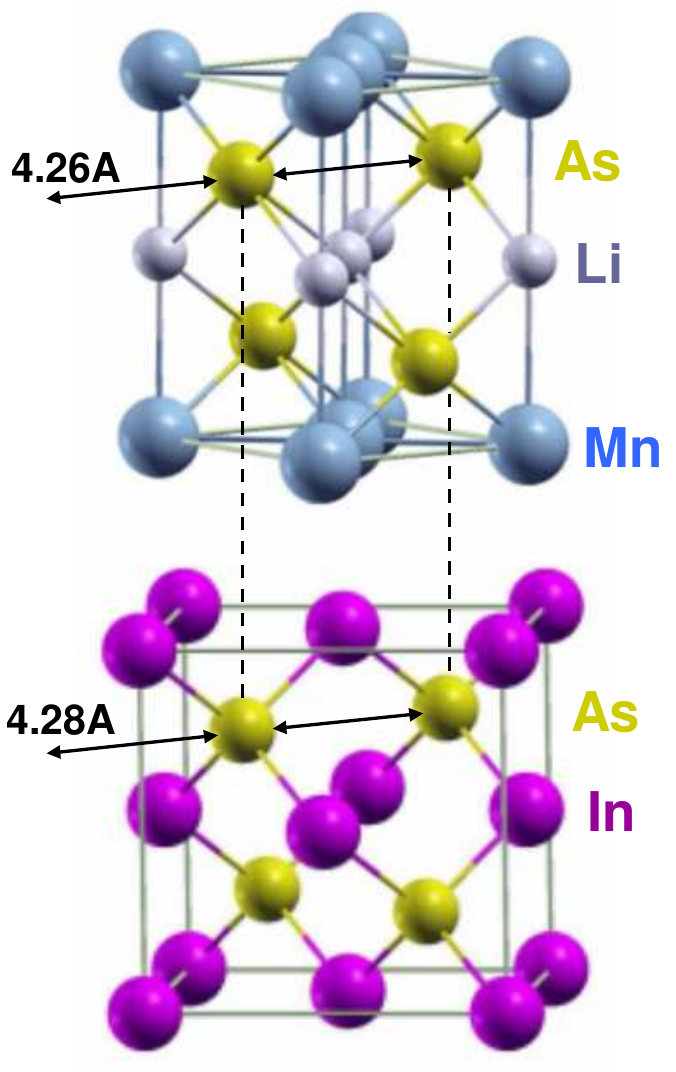}
\end{center}
\caption{Crystal structures of LiMnAs (top) and InAs (bottom). The As-sublattices almost perfectly match when the LiMnAs primitive cell is rotated by 45 degree against that of InAs.
}
\label{latt}
\end{figure}

The zinc-blend InAs with the cube lattice parameter 6.06~\AA~and the tetragonal LiMnAs have identical arrangements in the lattice of the common As planes and the respective As-As bonds are of a very similar length (the difference is 0.4\%), as illustrated in Fig.\ref{latt}. Therefore from the lattice matching perspective, InAs(100) is a particularly suitable common substrate material for the epitaxial growth of LiMnAs. The InAs substrate was mounted in the In-free sample holder, heated up to 450$^\circ$C for the oxide desorption, and coolled down to the growth temperature of 200$^\circ$C (measured by the thermocouple).

\begin{table}
\begin{tabular}{c | c c c c c c}
source & \,\, $T (^\circ$C) \,\, &  \,\, BEP  (Torr) \,\,  & \,\,  $\Phi/\Phi_{\rm Mn}$ \,\, \\
\hline
As & 260 & 5.7 $\times 10^{-8}$ & 1.1 \\
Li & 430 & 3.0 $\times 10^{-9}$ & 1.0 \\
Mn & 840 & 1.3 $\times 10^{-8}$ & 1.0 \\
\end{tabular}
\caption{Parameters of the MBE sources used during the growth of LiMnAs. The ratio of the elemental fluxes $\Phi$ was
computed from the beam-equivalent pressure (BEP) according to \cite{Parker}. 
}
\label{tab1}
\end{table}

\begin{figure}[h]
\vspace*{0.3cm}
\begin{center}
\includegraphics[width=0.5\columnwidth]{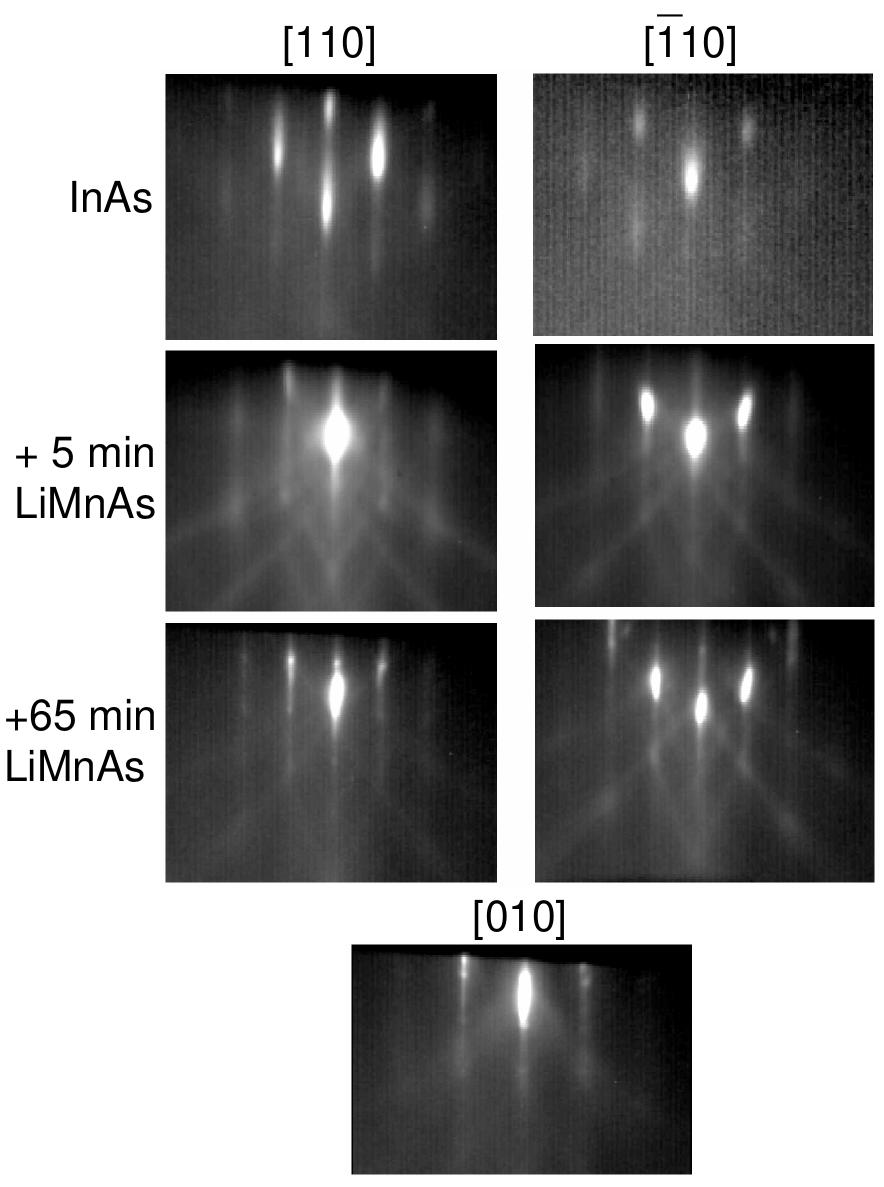}
\end{center}
\caption{RHEED image of InAs substrate before the growth (top panel), and during the growth of the LiMnAs film. The indicated directions refer to the InAs lattice.
}
\label{LiMnAs_rheed}
\end{figure}

We started the growth of LiMnAs directly on the desorbed InAs substrate surface, without any MBE grown buffer layer. Very quickly after opening the cell shutters a clear, streaky RHEED pattern evolves, showing the (1x1) symmetry of the surface and the 2D growth mode, as ilustrated in Fig.~\ref{LiMnAs_rheed}. This RHEED pattern remained stable even throughout the longest growth time attempted in our experiments which was three hours, corresponding to the film thickness of $\approx 0.6\,\mu$m. 

The RHEED images of growing LiMnAs are in a striking contrast to those recorded during the growth of a reference MnAs epilayer. If  the Li-cell shutter is closed during the growth, under otherwise identical growth conditions, a spotty RHEED pattern quickly emerges which corresponds to the mismatched and poorly growing hexagonal lattice of MnAs \cite{Tanaka}.

\begin{figure}[h]
\vspace*{0.3cm}
\begin{center}
\includegraphics[width=0.5\columnwidth]{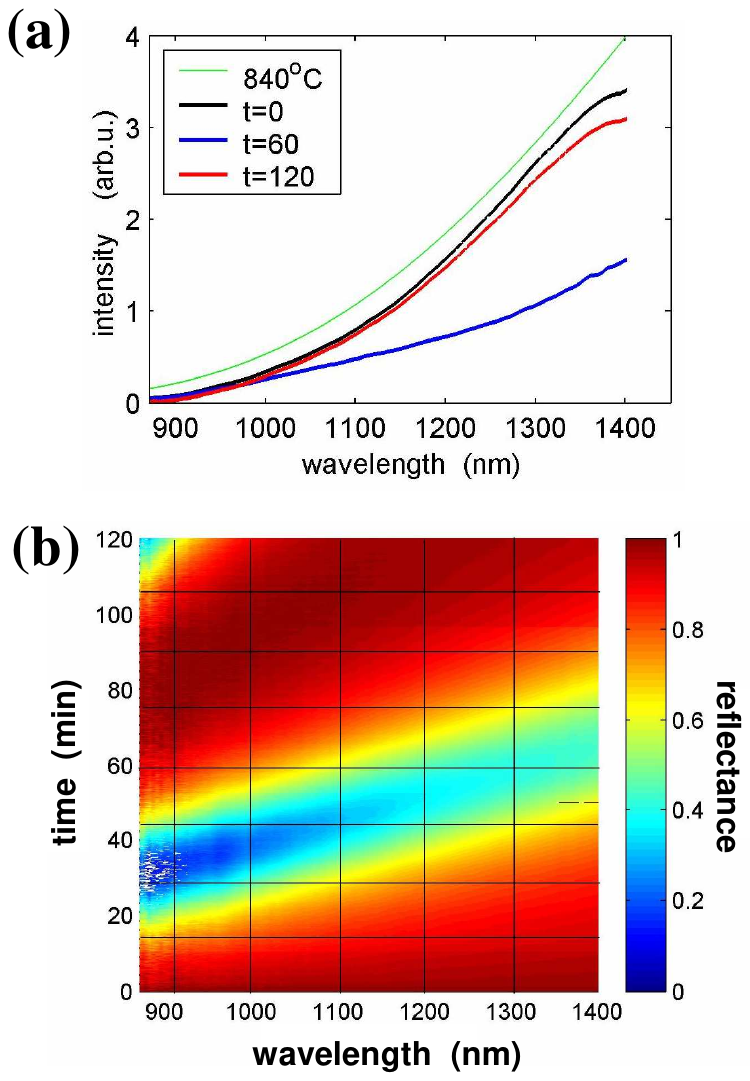}
\end{center}
\caption{(a) Spectra of the radiation reflected by the growing layer at the beginning of the growth (black curve),
and after 60 and 120 minutes of the growth (blue and red curve, respectively). For comparison, the thin green 
line corresponds to the radiation spectrum of a black-body at 840$^\circ$C ($\approx$ Mn cell). (b) Time evolution of the reflectance spectrum of the growing layer; at each time the reflectance was obtained from the spectrum of the reflected radiation
by dividing it by the spectrum at time $t=0$.
}
\label{specs}
\end{figure}

The RHEED growth oscillations were not observed during the growth of LiMnAs. Instead, we calibrated the growth rate ex-situ by measuring (by the Dektak profilometer) the total film thickness across the wafer edges masked during the growth by the sample holder. The reflectance spectroscopy was employed for monitoring the growing LiMnAs layer in-situ. For this purpose we use  the radiation which is emitted by the opened hot cells and reflected back by the growing layer. The reflected radiation is recorded by the spectrometer (kSA BandiT, detection range 870-1400~nm) mounted on the central pyrometer port. At the beginning of the growth the detected spectrum closely resembles that of the black-body radiation with temperature close to the temperature of the Mn cell, as shown in Fig.~\ref{specs}(a). During the growth the reflected intensity exhibits clear interference oscillations with the wavelength dependent period. The corresponding  time-dependent dimensionless reflectance is shown in Fig.~\ref{specs}(b). The occurrence of Fabry-Perot interference oscillations implies that  the growing film is transparent in the detected spectral range. 

\begin{figure}[h]
\vspace*{0.3cm}
\begin{center}
\includegraphics[width=0.45\columnwidth]{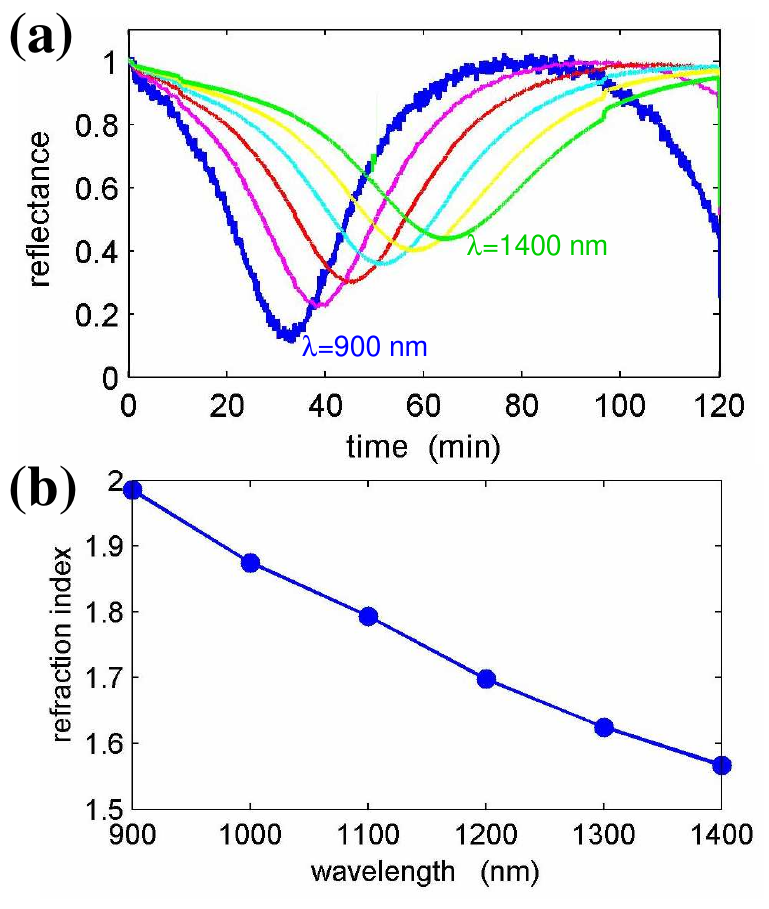}
\end{center}
\caption{(a) Reflectance oscillations recorded at wavelengths of 900~nm (blue curve, shortest period), 1000, 1100, 1200, 1300, and 1400~nm (green curve, longest period). (b) Wavelength dependence of the refraction index computed from the first half-period of the interference oscillations.
}
\label{osci}
\end{figure}

If combined with the known growth rate, the interference period allows us to estimate the index of refraction, as shown in Fig.~\ref{osci}. The resulting value of $n \approx 2$ is relatively low compared to common $sp$-band semiconductors.  Note that theoretically the low refractive index of LiMnAs is explained by means of the admixture of the correlated Mn $d$-states in the density of states \cite{Jungwirth:2010}.

\begin{figure}[h]
\vspace*{0.3cm}
\begin{center}
\includegraphics[width=0.6\columnwidth]{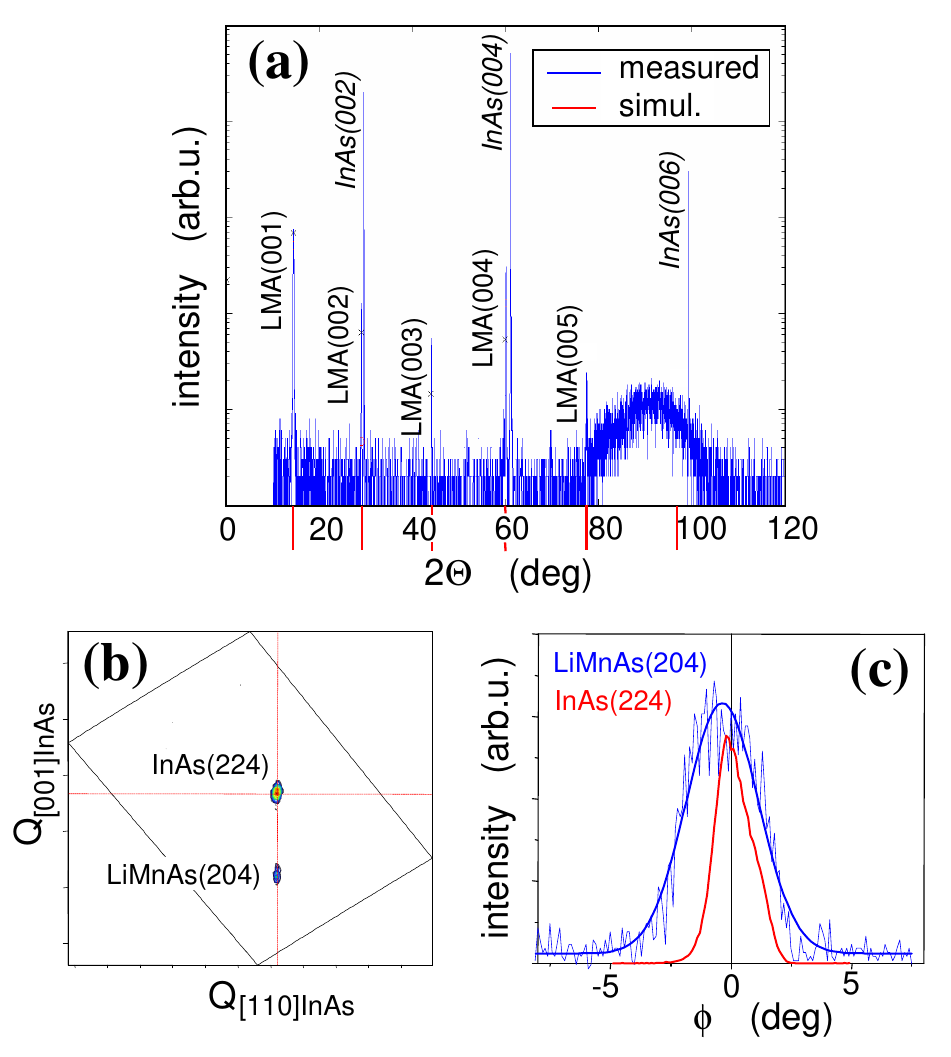}
\end{center}
\caption{(a) $2\Theta$-$\omega$ scan of LiMnAs; the peaks attributed to LiMnAs are denoted as LMA. (b) Reciprocal space map around the InAs(224) and LiMnAs(204) reflections. (c) $\phi$-scan around the LiMnAs(204) and InAs(224) using the $2\Theta$ and $ \omega$ coordinates determined from (b).
}
\label{xrd}
\end{figure}

LiMnAs samples of thickness 200~nm were analyzed by XRD. The $2\Theta$-$\omega$ scan in Fig.~\ref{xrd}(a) shows peaks from the (001) family. Red lines denote the expected bulk positions for LiMnAs(001) reflections which are in a perfect agreement with the observed peaks. The bump at 90$^\circ$ may correspond to a small amount of spurious contamination. The epitaxial layer is fully strained, as seen from Fig.~\ref{xrd}(b). We verified by $\phi$-scan around LiMnAs(204) and InAs(224) reflections (see Fig.~\ref{xrd}(c)) that the epitaxial relationship is $[100]$LiMnAs(001)$\parallel [110]$InAs(001) in agreement with the expected relative crystal orientation of LiMnAs grown on InAs  (see Fig.~\ref{latt}).

\section{Conclusions}

We have prepared high quality, single crystal thin film samples of LiMnAs by the standard solid source MBE technology which demonstrates the feasibility of epitaxial growth of group-I alkali-metal compounds. The LiMnAs epilayers have the expected crystal structure and epitaxial relationship to the InAs substrate. The materials exhibit an optical gap. Our work opens the possibility for a systematic study of material, optical, electronic, and magnetic characteristics of the family of I-Mn-V antiferromagnetic semiconductors.

\bigskip

\section*{Acknowledgments}
We greatfully acknowledge helpful assistance of J.~Ma\v{s}ek, P.~Ku\v{z}el, B.~L.~Gallagher,
C.~T.~Foxon, R.~P.~Campion, and J.~Wunderlich. The work was done in the framework of 
EU Grant FP7-214499 NAMASTE, FP7-215368 SemiSpinNet, from Czech Republic Grants AV0Z10100520, 
AV0Z10100521, MSM0021620834, MSM0021620857, KAN400100652, LC510, and Preamium Academiae.

\end{document}